\documentclass[conference]{IEEEtran}
\IEEEoverridecommandlockouts
\usepackage{cite}
\usepackage{amsmath,amssymb,amsfonts}
\usepackage{algorithmic}
\usepackage{graphicx}
\usepackage{textcomp}
\usepackage{xcolor}
\usepackage{caption}
\usepackage{subcaption}
\usepackage{hyperref}
\usepackage[utf8]{inputenc}
\usepackage[T1]{fontenc}
\usepackage{amsthm}
\usepackage{multirow}
\usepackage{fontawesome}
\usepackage{enumitem}
\usepackage{footmisc}
\usepackage{float}
\usepackage{url}
\usepackage{soul}
\usepackage{tikz,pgfplots}
\usepackage{tikz-qtree}
\usepackage{algorithm}
\usetikzlibrary{trees}
\usepackage{verbatimbox}
\usepackage{listings}
\usepackage{mathtools} 

\tikzstyle{selected}=[draw,minimum width=1in,text width=2.5in, align=center, anchor=west]

\usepackage[export]{adjustbox}
\pgfplotsset{compat=newest}
\definecolor{mygr}{HTML}{e6e6e6}

\theoremstyle{definition}

\def\BibTeX{{\rm B\kern-.05em{\sc i\kern-.025em b}\kern-.08em
    T\kern-.1667em\lower.7ex\hbox{E}\kern-.125emX}}

\begin{document}

\title{VCTP: A Verifiable Credential-based Trust Propagation Protocol for Personal Issuers in Self-Sovereign Identity Platforms}

\author{
    \IEEEauthorblockN{Rahma Mukta\IEEEauthorrefmark{1}, Rue C. Teh\IEEEauthorrefmark{1}, Hye-young Paik\IEEEauthorrefmark{1}, Qinghua Lu\IEEEauthorrefmark{2} and Salil S. Kanhere\IEEEauthorrefmark{1}}
    \IEEEauthorblockA{\IEEEauthorrefmark{1}School of Computer Science and Engineering, UNSW Sydney
    \\\{r.mukta, r.teh, h.paik, salil.kanhere\}@unsw.edu.au}
    \IEEEauthorblockA{\IEEEauthorrefmark{2}Data61, CSIRO, Sydney,
    qinghua.lu@data61.csiro.au}
}


\maketitle

\begin{abstract}
Self Sovereign Identity (SSI) is an emerging identity system that facilitates secure credential issuance and verification without placing trust in any centralised authority. To bypass central trust, most SSI implementations place blockchain as a trusted mediator by placing credential transactions on-chain. Yet, existing SSI platforms face trust issues as all credential issuers in SSI are not supported with adequate trust. Current SSI solutions provide trust support to the officiated issuers (e.g., government agencies), who must follow a precise process to assess their credentials. However, there is no structured trust support for individuals of SSI who may attempt to issue a credential (e.g., letter of consent) in the context of business processes. Therefore, some risk-averse verifiers in the system may not accept the credentials from individual issuers to avoid carrying the cost of mishaps from potentially inadmissible credentials without reliance on a trusted agency. This paper proposes a trust propagation protocol that supports individual users to be trusted as verifiable issuers in the SSI platform by establishing a trust propagation credential template in the blockchain. Our approach utilises (i) the sanitizable signature scheme to propagate the required trust to an individual issuer, (ii) a voting mechanism to minimises the possibility of collusion. Our implementation demonstrates that the solution is both practical and performs well under varying system loads.
\end{abstract}


\section{Introduction}
\label{intro}

Self-Sovereign Identity (SSI), as a decentralised identity management system, is a privacy-friendly alternative to centralised and proprietary \textit{Id}entity \textit{M}anagement \textit{S}ystems. Many SSI platforms  employ blockchain to leverage decentralisation for identity credential issuance and verification. In SSI, issuers of credentials play a central role. An issuer (e.g., administration office of University A) creates a Verifiable Credential (VC)\footnote{We use the terms ``verifiable credentials'' and ``credentials'' interchangeably. } (e.g., an academic transcript) and links it with a DID (Decentralised IDentifier) of the credential holder (e.g. Jane, a student of University A), who then presents it to a verifier (e.g., a potential employer of Jane). To accept the identity presented in a VC, a verifier must trust the issuer of the VC.  

In current SSI platforms \cite{b1, b2}, verifiers are usually aware of the (public) DIDs of officiated credential issuers (e.g., accredited universities and government offices). However, non-officiated issuers, i.e., individual users in SSI, are not afforded the same level of trust within the framework, although they can issue ``personally'' signed credentials. From here on, we refer to the non-officiated issuers as \textit{personal issuers}. Some risk-averse verifiers in the system may shun the credentials from the personal issuers to avoid carrying the cost of mishaps from potentially inadmissible credentials in their business processes \cite{b3}. Given that most business processes are driven by users submitting documents or filling forms, overlooking the trust of personal issuers could limit the scope of SSI applications. Therefore, this paper aims to add trust to the personal issuers.

As a motivating scenario, let us suppose that \textbf{Anna} has an accident at work and is admitted to a hospital for an emergency operation. Anna wants \textbf{Tim}, her friend, to be able to sign some hospital admission forms for her.  She is told by \textbf{the treating doctor} that \textbf{the hospital policy} requires Anna to sign a ``letter of authority''. In this scenario, we consider Anna as the personal issuer who signs/issues the letter (credential) for Tim. Later in the hospital, Tim, the holder of the letter, presents it to \textbf{a physiotherapist}. Besides verifying Tim's identity, the physiotherapist also wants to confirm that the signature is from Anna and that Anna can issue such a credential. 

In this scenario,  the treating doctor could attest to Anna's identity and support her status as an issuer (i.e., give credence to her as a legitimate issuer of the letter). We refer to this process as ``personal issuer onboarding'' whereby an existing issuer with appropriate authority can  \textit{support} a personal issuer. We assume that a credential, such as the letter of authority, will be created as a template in which necessary data fields and their usage rules are defined. In fact, many business forms share the same notion. In SSI, such a template could be managed by an officiated issuer (Level 1, e.g., the hospital). The officiated issuer also ``seals'' this template by signing it. This ensures the verifier can verify if the letter is correct for the business process. 

In implementing such a process, there are critical considerations. First, when the letter is \textit{in use} (i.e., an instance of the letter template is created), its content needs to be updated (e.g., the treating doctor to add Anna into the letter as an issuer). Since it is sealed, these updates must happen without invalidating other sections of the letter (including any existing signatures before the update). Second, when the template is defined by the top level issuer, the users who may have to update the letter are not known. Third, when the trust is propagated to the next level issuer, we must minimise any possibility of collusion among the entities (e.g., a doctor onboarding an illegitimate person as a patient). 

Updating a credential, such as the letter in our scenario, is a challenge because an update should still guarantee the integrity and authenticity of the credential. Conventional digital signatures (e.g., RSA or DSA)\cite{b4} provide the means to achieve both credential integrity and issuer authentication. Other constructs, such as Redactable Signatures\cite{b1}, allow anyone to update a document without invalidating the existing signature from the original signer. However, the personal issuer onboarding process requires that only the duly authorised updater can update the document (e.g., treating doctors in the hospital) in a controlled and verifiable way in the {\it designated part} of the credential with information of the next level issuer. In this regard, our solution utilises the {\it Policy-based Sanitizable Signature Scheme} which allows a crytographical solution to open a backdoor to an already-signed credential so it can be updated by a designated updater without invalidating any existing signature on that credential. Combined with policy-based access control to such a backdoor, the scheme provides an ideal basis for building a protocol that adds trust to signers of a credential in a controlled manner.

In summary, we present the design and implementation of a verifiable credential based trust propagation (VCTP) protocol to add trust to personal issuers in an SSI-based ecosystem.   

\begin{itemize}
    
    \item We propose a design of a new type of \textit{verifiable credentials} where a credential may embed an update policy section prescribing for what purpose and by whom certain sections can be updated. We support the creation of such a credential by proposing a template named \textit{Trust Propagation Template}.
    
    \item We design and implement a series of cryptography algorithms that execute the update policy using the policy-based sanitizable signature. Although the signature scheme has been used in the context of mutable blockchain \cite{b5}, to the best of our knowledge, our work is the first in using such a method for issuer trust management in the SSI platform,
    
    \item A proof-of-concept of the protocol is implemented using health service consent delegation as an application scenario. Our implementation demonstrates that the solution is both practical and performs well under varying system loads. We also perform a qualitative security analysis of the proposed VCTP protocol where we consider several probable attack scenarios and argue how our protocol defends against them. 
    
\end{itemize}

The rest of the paper is organised as follows. 

Section \ref{sec:preli} presents the background concepts of the proposed VCTP protocol.  
Section \ref{sec:solution} describes the main workflows in VCTP protocol to onboard personal issuers in the SSI ecosystem.
Section \ref{sec:eval} presents the experimental analysis with our evaluation results.
Section \ref{sec:related-work} gives some background about issuer trust management and motivating our proposal, followed by a conclusion in Section \ref{sec:conclusion}.

\section{Preliminaries}
\label{sec:preli}

This section introduces the background concepts that support the VCTP protocol. Namely, an SSI platform, role credentials, the policy-based sanitizable signatures with its underlying construction techniques and a voting process to minimise potential issuer collusion.

\subsection{SSI Platform}

\begin{figure}[ht]
\centering
\includegraphics[width=0.95\linewidth]{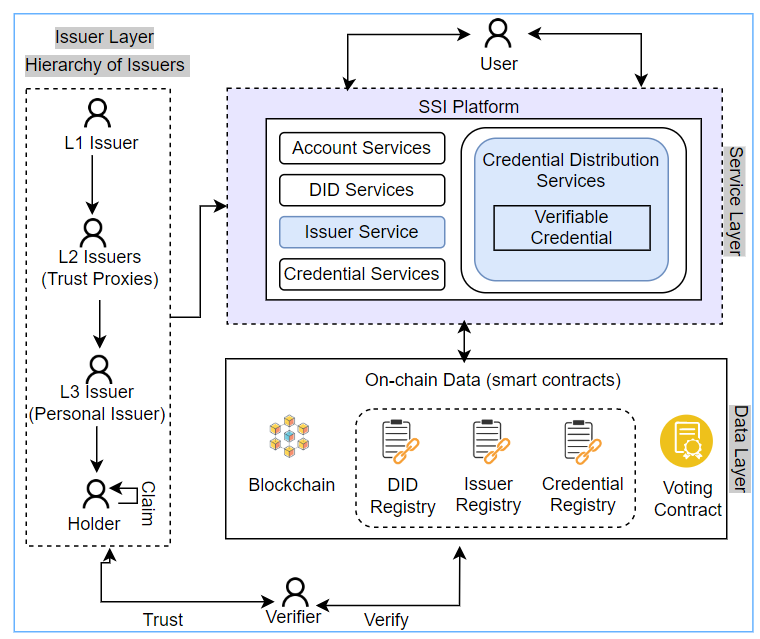}
\caption{System Components of SSI Platform with issuer hierarchy}
\label{fig:arch}
\end{figure}

Fig.~\ref{fig:arch} shows an SSI platform with levels of issuers\cite{b6}. For simplicity, we are showing three levels of issuers, i.e., L1, L2 and L3. We assume a governance authority is an external entity that sets the rules and regulations for all participating users. L1 issuers with the highest level of trust (e.g., academic institutions, public hospital administration) are endorsed by the governing authority according to governance policies. 
In this SSI platform, L1 issuers can endorse the next level issuers, who then can endorse the next level issuers and so on. This is done using a verifiable credential and the exact protocol described in this paper. However, for clarity, we omit the discussion on how the higher level issuers are endorsed and focus on the trust of personal issuers.
For this reason, in this paper, we refer to the issuers who are already in the system as \textit{trust proxies}.

The VCTP protocol relies on blockchain to operate registries that are relevant for managing the core SSI components such as DIDs (Decentralised Identifiers), credential records and issuers. 
The platform shown in Fig.~\ref{fig:arch} depicts the various components that implement SSI. The platform includes the service layer, which provides \textit{Account Services} (for blockchain account management), \textit{DID Services} (for managing DID accounts), \textit{Issuer Services} (for managing issuer eligibility to sign credential) and \textit{Credential Distribution Services} (for managing and verifying credentials). This layer abstracts and exposes all necessary functionalities to manage identities. The data layer includes on-chain data (the registry smart contracts). All the verified issuers’ DIDs are maintained in \textit{Issuer Registry}. To ensure identity data integrity, a \textit{DID Registry} is created as a smart contract on the blockchain, which maintains the mappings between registered DIDs and their associated DDOs. The hash of a signed credential is stored in the \textit{Credential Registry} as an issuance record. In addition, the voting process for each credential update is managed by the \textit{Voting Contract}.

\subsection{Role Credential (RC)}

This is a regular, verifiable credential that holds sets of attributes to identify a person depending on his position. These attributes, either individually or when combined, can attest to a membership of that person in a group or a unique identity that distinguishes that person from the others. For example, a doctor holds a doctor's registration number, name, position, address, contact, and birthdate as their attributes, whereas a student holds a student number, enrolment year, school, institution etc. These attributes are often called role attributes. These role attributes are typically captured in a role credential made by the authorised issuer, for example, by a hospital or educational institution, to authenticate the owner of the role credential in later communications.

\subsection{Sanitizable Signatures and Related Functions}
\label{subsec:PCH}

Sanitizable signatures allow a signer to partly delegate signing rights to a semi-trusted party, called a sanitizer (we refer to it as ``updater'' from here on to make the term more descriptive for our protocol). Sanitizable signatures introduced in \cite{b7}, presented its generic construction based where a message (e.g., a credential in this paper) is first hashed using  Chameleon Hash (CH), then signed by the root signer (e.g., Level 1 issuer in our scenario). When the CH technique is applied, the root signer can choose a specific updater who can later update predetermined sections of the credential while  maintaining the same originally signed hash. 

\textbf{Chameleon Hash Functions \cite{b7,b8}} enable this feature by incorporating a trapdoor mechanism. A party who is privy to the trapdoor can find random collisions in the domain of the functions. That is, such a party can update  a credential and still keep the original hash of the credential unchanged after each update. 

To explain the function further, a chameleon hash is defined by the triplet: $\mathtt{(Gen, CH, CH^{-1})}$, where $\mathtt{Gen}$ is a key generation algorithm that generates a key pair $\mathtt{(hk,td)}$. $\mathtt{hk}$ is the hashing (public) key and $\mathtt{td}$ being the trapdoor (private) key. A chameleon hash function $\mathtt{CH()}$ is given $\mathtt{hk}$, a credential $C$ and a random string $\mathtt{r}$, then generates a hash value $\mathtt{CH_{(C, r)}}$. $\mathtt{CH()^{-1}}$ is the corresponding reverse function that uses $\mathtt{td}$. Given the pair $\mathtt{CH_{(C, r)}}$ and any updated credential $C'$, the reverse function outputs a value $\mathtt{r'}$ such that $\mathtt{CH_{(C, r)} = CH_{(C', r')}}$. It must be noted that the chameleon-hash functions are collision resistant as long as the corresponding trapdoor key $\mathtt{td}$ is not known.

\textbf{Managing the trapdoor keys} becomes a critical issue in applying CH, both from a security viewpoint (anyone with a trapdoor key will be allowed to update) and from an operational viewpoint (how to manage who should have the key).
A solution is adding an access control policy to the trapdoor key via attribute-based encryption (ABE) techniques. With an access policy in place, an updater must satisfy predefined access attributes to be given the key.

\subsection{Policy-based Sanitizable Signatures}
\label{subsec:CH}

A sanitizable signature scheme using Chameleon Hash, whose trapdoor keys are managed with an access policy, is called \textit{Policy-based Sanitizable Signatures}. When computing a credential hash, the public keys of required access attributes are included so that only entities who possess private keys to the corresponding attributes can obtain the trapdoor key. In generic construction, Policy-based Sanitizable Signatures are a combination of the concept of CH \cite{b8} and Attribute-Based Encryption (ABE)\cite{b9} scheme.

We use this scheme to allow an L1 issuer to (i) specify updatable sections of the credential, (ii) for each updatable section which has its own trapdoor key, who (in terms of attributes) can access the key. Each updatable section is individually hashed using the access policy-based chameleon hash functions. All sections are then combined and signed. Thus, the implemented signature is defined as $\mathtt{\sigma = SIGN(PCH_{(s_1)}||...||PCH_{(s_n))}}$, where $\mathtt{PCH_{(s_i)}}$ represents the chameleon hash of each section. To be precise, in our implementation, each trapdoor key $td_i$ is encrypted using ABE. An updater can update the credential, if, and only if, it holds a valid $td_i$ and the private keys to access attributes.

\subsection{Voting Process}

Although a cryptographical scheme such as sanitizable signature can ensure only the designated updater can update credentials, the scheme alone cannot mitigate the challenges arising from human users colluding to onboard illegitimate users. For example, a malicious updater may collude with a malicious external user and try to update a credential by adding attributes of the malicious user as the personal issuer. Such a security issue has been studied in existing literature\cite{b10, b11} where participating entities are asked to vote on choices to compute the trust value of untrusted users. Our voting process is designed in a way that voters are asked to review someone's action (e.g., adding Anna as a patient) and vote to confirm  if the action is legitimate. We believe this ``human review'' based process can be used in conjunction with other collusion detection algorithms to further minimise the collusion in the system.

Since our SSI platform is blockhain-based, we surveyed a series of voting patterns on a blockchain platform \cite{b12}. We adopt one of the patterns to design the voting process for our protocol. In our voting process, the voting requires a voter to reviews some content (i.e., the proposed update of a credential by a trust proxy). The voting action practically indicates if the voter is satisfied that the credential update is legitimate.

\section{An Overview of VCTP Protocol}
\label{sec:solution}

This section presents the overall design of the VCTP protocol which consists of two main components: credential templates for issuer trust propagation, and a voting mechanism for minimising potential issuer collusion.

\subsection{Trust Propagation Templates}

In this section, we introduce \textit{trust propagation templates} in detail. In conventional SSI systems, individuals use credentials to generate claims on certain identity properties about themselves. A trust propagation template is a type of credential that encodes the credential update policy by which the trust of higher level issuers is propagated through the trust proxies to personal issuers. 
Listing~\ref{listing:SP} shows an example of a trust propagation template embedded in a ``letter of authority'' as a credential. A template is divided into sections as follows:

The \textbf{Update Policy} section is defined by the L1 issuer and contains the rules and usage conditions.
 
        \begin{itemize}
            \item \textit{context}, the URL where the human-readable attribute terms of the template are defined
            \item \textit{id}, the identifier of the credential
            \item \textit{jurisdiction}, the domain in which this template is valid,
            \item \textit{type}, sets the context of the credential as ``trust propagation''. 
            \item \textit{officialIssuer}, contains the DID of the issuer who defines the policy  
            \item \textit{issuanceDate}, the issuance date of the template 
            \item \textit{expirationDate}, the expiration date of the template
            
            \item \textit{policy}, sets (i) the attributes of trust proxies (e.g., \textit{treating doctor}) with permissible actions by the proxy, and (ii) the attributes of the next level issuer. Note that in our scenario, these will be personal issuers. 
    \end{itemize}
    
        Further, the following fields configure how the voting process will be run. 
\begin{itemize}
    \item \textit{scenario}, dictating if any update attempt on the credential template requires voting to be undertaken.
    \item \textit{approvalPolicy}, specifying who can vote via the attributes from a role credential (RC).
    \item \textit{numVotesRequired}, specifying the number of votes required to satisfactorily complete the voting process (described in Section~\ref{subsec:voting}).
\end{itemize}

The \textbf{Trust Proxy} section is to be updated by designated trust proxies. It contains the DID of the trust proxy, and the DID of the personal issuer the trust proxy is onboarding (e.g., in our scenario, Anna the patient).
    
The \textbf{Credential} section shows the  credential itself with the title and relevant text. The designated personal issuer signs and issues it to the intended holder (e.g., Tim in our scenario).

\begin{lstlisting}[language=html, label={listing:SP}, caption=An example of Trust Propagate Template embedded in a credential ``Letter of authority'', basicstyle=\ttfamily\footnotesize]
[{ // update policy
"@context":https://www.w3.org/2018/credentials/v1,
"id": "http://example.edu/credentials/1872",
"jurisdiction": "HospitalA",
"type": ["VerifiableCredential","TrustPropagation"],
"officialIssuer": "did:example_hos:fcgfc2g823fcdd387",
"issuanceDate": "2021-07-10T04:20:00Z", 
"expirationDate": "2021-07-17T04:20:00Z",
"scenario": "OutPatient",
"approvalPolicy": ["doctor", "nurse"],
"numVotesRequired": 5,
"policy": {
   { "proxyAttribute": ["doctor", "HospitalA"],
     "permissions": ["propagate-trust"] },
     "nextLevelIssuerAttrs": ["patient", "HospitalA"] 
   }
},
{ // trust proxies
"TrustProxy" : "did:example_doctor:fcgfc2g823fcdd387",
"nextLevelIssuerDetails": {
"id": "did:example_patient:fcgfc2g823fcdd387",
"permissions": ["delegate-medical-decision"]}
},
{ // credential itself
"Title":Letter of Authority,
"IssueDate":2/2/2022,
"Text":this letter is to authorise the person named in
the document to act on my behalf in matters related 
to the subject mentioned in the document.}
{ 
"signedBy":"did:example_patient:fcgfc2g823fcdd387",
"credentialSubject":{ // holder
"id":"did:example_holder:fcgfc2g823fcdd387",
"permissions":["routine-medical-care"]}
}]
\end{lstlisting}

\subsection{Trust Propagation using Policy-based Sanitizable Signatures}
\label{subsec:TP-PSS}

\begin{figure}[ht]
\centering
\includegraphics[width=0.99\linewidth]{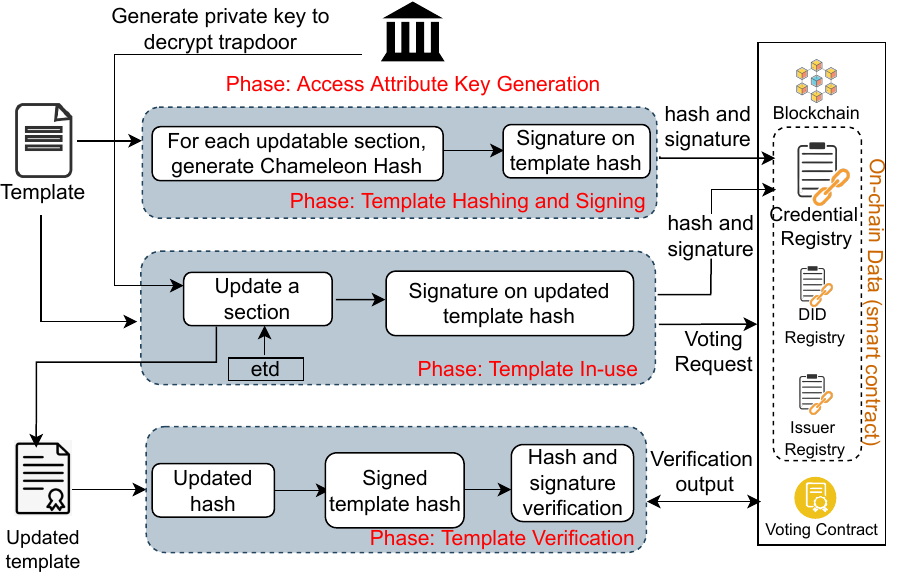}
\caption{Issuer trust propagation using policy-based sanitizable signatures}
\label{fig:chain}
\end{figure}

Now we describe the policy-based sanitizable signature scheme which executes the credential update policy. The scheme consists of five algorithms, each implementing a key step (e.g., generating keys, hashing credentials) in the workflow. We illustrate the overall workflow of the scheme in Fig.~\ref{fig:chain} over four distinct phases, namely: Access attribute key generation, Template hashing and signing, Template in-use and Template verification.

\paragraph{\textbf{Phase: Access Attribute Key Generation}} L1 issuers are in charge of managing the required access attributes in the system. In this phase, they use two key generation algorithms to generate a pair of public/private hashing keys and a master key for each access attribute. In our implementation, we assume that this step is done once by an L1 issuer at the start of an SSI application (e.g., a hospital credential management). 

\paragraph{\textbf{Phase: Template Hashing and Signing}} The main algorithm used in this phase is named $\mathtt{Hash_{PCH}()}$. Given a trust propagation template ($\mathtt{T}$), $\mathtt{Hash_{PCH}()}$ works as follows:  
(1) picks each updatable section $\mathtt{s_i}$ of $\mathtt{T}$, and generates its chameleon hash $\mathtt{CH_i}$ and attribute-encrypted trapdoor $\mathtt{etd_i}$ (the access attributes for each trapdoor is defined by L1 issuer), (2) All $\mathtt{CH_i}$ are concatenated and combined with the remaining sections of $\mathtt{T}$. Then the algorithm computes the hash of $\mathtt{T}$ and signs it with the key of the relevant L1 issuer. 
The hashed credential along with the signature $\mathtt{\sigma}$ are stored in the \textit{credential registry} (shown in Fig.~\ref{fig:arch}) for credential integrity and future verification.

\paragraph{\textbf{Phase: Template In-Use}} 
In this phase, trust proxies update the sections of the template to onboard the next level issuers. It is noted that in our current implementation, we assume each authorised updater receives the following three items through a secure channel; the correct hashed template, public hashing key for generating a chameleon hash for their authorised section ($\mathtt{s}$) of the template, and the attribute-encrypted trapdoor $\mathtt{etd}$. 

The main algorithm used in this phase is named $\mathtt{Update_{PCH}()}$. Each updater needs to complete the following three steps using this algorithm:
\begin{itemize}
    \item Firstly, to decrypt $\mathtt{etd}$, the updater must obtain the private keys of the attributes with which $\mathtt{etd}$ is encrypted. The updater sends the required attributes to the access attribute key generator. This can be done by disclosing relevant identity credentials. Upon successful attestation of the attributes, the updater receives the private keys of the corresponding attributes. 
    
    \item Secondly, after decrypting $\mathtt{etd}$, the updater updates the section (say $\mathtt{{s'}}$ and recomputes new randomness $\mathtt{r'}$ such that $\mathtt{hash(s, r) = hash({s'}, {r'})}$ (described in \ref{subsec:PCH}). Thus the hash of the updated section is not altered, only the random string is altered. 
    
    \item Thirdly, the updater sends the hash of the resulting credential to \textit{credential registry} as a transaction. It is noted that each updater signs the updated section with their own key, and the hash of the resulting credential, of course, is the same as the original template because of the trapdoor and chameleon hash functions. The signatures of the updaters can be used to ensure accountability when necessary.

\end{itemize}

The update policy of the template may require voting to take place before the credential is committed to the registry. This voting process allows other stakeholders in the business system to ensure that the credential update is appropriate, i.e., an issuer is not intentionally defining invalid permissions in the credential or trying to onboard a fake issuer. The voting process is described below in Section~\ref{subsec:voting}.

\paragraph{\textbf{Phase: Template Verification}}

In this phase, the credential is presented for a verification (e.g., Tim presenting the letter to the physiotherapist). It includes the hash of the credential and its corresponding signature $\mathtt{\sigma}$. Verifiers can verify them from \textit{credential registry} using verification algorithm named $\mathtt{Verify_{PCH}()}$. This deterministic algorithm gets as input the public hashing key, a credential section $\mathtt{s}$, randomness $r$, and the hash of section $s$. It outputs a decision $\mathtt{d \epsilon (0, 1)}$ indicating whether the hash is the valid hash of $s$. This verification further ensures that the credential is the correct template issued by L1 issuer. If necessary, the verifiers can also access corresponding DID information (i.e., public key of credential holder and issuer) from the \textit{DID registry} and \textit{issuer registry} (shown in Fig.~\ref{fig:arch}).

\subsection{Voting Process}
\label{subsec:voting}

\begin{figure}[ht]
\centering
\includegraphics[width=0.99\linewidth]{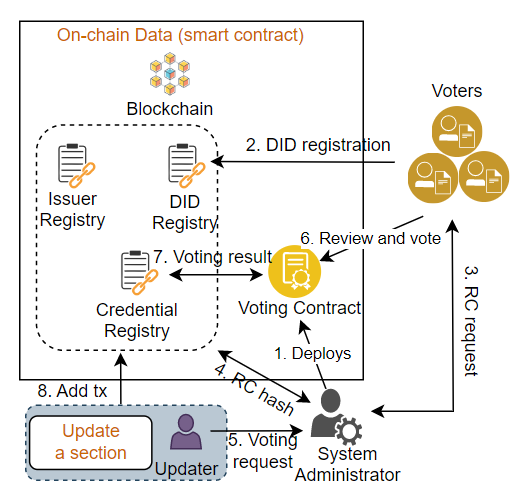}
\caption{Voting Process in VCTP}
\label{fig:voting}
\end{figure}

Our voting process includes three phases: (1) system initialisation, (2) voting update, and (3) voting management shown in Fig.~\ref{fig:voting}.

\subsubsection{System Initialisation} This phase (Steps~(1-4) in Fig.~\ref{fig:voting}) essentially initialises the system by configuring a group of parameters and completing key generations for the voting process. L1 issuer defines the number of votes required (specified in the ``numVotesRequired'' field shown in example template), as well as the roles of the voters (specified in the ``approvalPolicy'' field shown in the example template) that can ``vote'' for a credential update within the organisation.

\paragraph{\textbf{System Setup}} 

A System administrator, who takes the responsibility to be a voting administrator, generates its own key pair(s) for the deployment of the \textit{voting contract} and issuance of role credentials to the new voters.

\paragraph{\textbf{Registration}} Each voting entity must have a DID and role credential (e.g., as a doctor or nurse of a hospital) as an internal entity (or employee) of the organisation (e.g., a hospital). The role credential should be associated with DID to prove the voter's identity and voting right.

\subsubsection{Voting Update} This phase addresses major operations of the voting implementation in a smart contract, including voting request, vote, and count as shown in Steps~(5-7) in Fig.~\ref{fig:voting}.

\paragraph{\textbf{Voting Request}} When an updater attempts to commit the updated template to the \textit{credential registry}, it is first decrypted to determine if the voting process is required. If required, the voting request is sent by creating a new entry in the \textit{voting registry}.

\paragraph{\textbf{Review and Vote}} Each voter (e.g., doctors, nurses and receptionists associated with Anna's treatment process) reviews the update made in the template and votes to confirm the correctness of updated information (e.g., the voters will identify Anna as the ``valid patient'' who is going to be a personal issuer). The voter encrypts his signature (generated using the voter's private key from DID) with his vote option by using the voting contract’s public key and sends them to the contract.

\paragraph{\textbf{Voting result}} The voter's signature is verifiable using the voter's public key (associated with the voter's DID from role credential). The \textit{voting contract} verifies the signature and accepts the vote if the voter has an appropriate role, and it is not a duplicated vote from the same DID. Finally, the voting contract computes the voting result.

\subsubsection{Voting Management} In this phase at Step~8 in Fig.~\ref{fig:voting}, before adding a transaction (\textit{tx}) on-chain coming from the updater is first verified against the voting result calculated at \textit{voting contract}. If the result satisfies the number of votes required for the \textit{tx} to be committed, then \textit{tx} (i.e., hash(es) of the updated section(s) and the signature is added to the \textit{credential registry}. Finally, the associated credential subject from the updated section is added as an issuer in the \textit{issuer registry}. Otherwise, if the voting result does not satisfy the number of votes required, the \textit{tx} will be rejected by the \textit{credential registry} and \textit{issuer registry} which will leave the issuer unverifiable on-chain.

\section{System Evaluation}
\label{sec:eval}

To evaluate the proposed VCTP protocol, we first demonstrate the feasibility by implementing a concrete instance of the protocol. Second, we benchmark its performance with system load tests and evaluate the performance of the voting process. Finally, we discuss the security properties of the protocol.

\subsection{Implementation}

\begin{figure}[ht]
\centering
\includegraphics[width=0.99\linewidth]{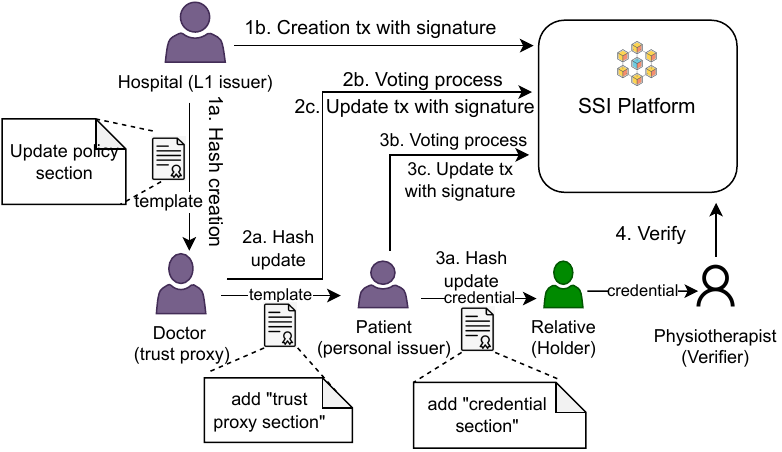}
\caption{End-to-end Testing Scenario for Analysis}
\label{fig:test}
\end{figure}

We implemented the proposed protocol with all the workflows and blockchain registries described in Section~\ref{sec:solution}. Fig.~\ref{fig:test} demonstrates our end-to-end testing scenario in a healthcare context. (1) Hospital, in the L1 issuer role, generates the template to facilitate the update process (e.g., letter of authority). (2) A Doctor (trust proxy) instantiates the template as a credential and updates it to onboard his patient (personal issuer) and then (3) patient updates the credential to issue it to his relative. Finally, the relative presents the credential to a physiotherapist regarding an emergency treatment option. (4) The physiotherapist, acting as a verifier, verifies the credential from SSI platform.

Our implementation is based on Python and uses Linux virtual machine to conduct the analysis. The SSI platform services and registry smart contracts are deployed on a 1.8 GHz Intel Core i5 machine with 16GB RAM. All entries to the blockchain (such as DID registration) occur through smart contracts that are written in Solidity. The deployment of smart contracts utilises \textit{brownie}\cite{b13}, a python-based development and testing framework using the Ethereum Virtual Machine. We undertake the unit tests for smart contracts (e.g., \textit{Voting Registry}) using the \textit{brownie} package to ensure that functions work as intended by comparing the actual and expected outputs. In addition, an end-to-end test is performed to ensure that system components are integrated.

\subsection{Performance Analysis}

We deployed the blockchain node with the registry smart contracts on a Linux Virtual Machine with 4 cores and 16GB RAM in Microsoft Azure. The client requests were generated on a local virtual machine with 2 cores and 8GB RAM. We performed tests to understand the overhead of template hash creation, update and verification process (described in Section~\ref{subsec:TP-PSS} as \textit{phase: Template hashing and signing}, \textit{phase: Template in-use} and \textit{phase: Template verification}). Next, to understand the blockchain performance we tested the template hashing and signing as well as update processes with the associated blockchain \textit{tx} generation time. We did not consider template verification here as this does not generate any new \textit{tx} on-chain. Finally, we analysed the voting process to understand the time cost while minimising the probability of collusion. This is to be noted here, we did not compare the proposed VCTP protocol with a baseline system. Any SSI platform which allows registered issuers \cite{b1, b2} can be considered as a baseline system of our work. However, these baseline systems did not add the additional steps that we have proposed while propagating trust to the issuers. Without such steps comparing our work with a baseline system may not sound logical.

\paragraph{\underline{Performance of Template Hashing Algorithms}} 

We consider the execution time of template hash creation ($\mathtt{Hash_{PCH}()}$), update ($\mathtt{Update_{PCH}()}$) and verification($\mathtt{Verify_{PCH}()}$) over 100 runs. We increased the number of attributes from 4 to 32 attributes in a template, because with an increasing number of issuer levels, the number of attributes may increase. From Fig.~\ref{fig:ExecTime} it is worth noting that template hash creation requires 0.9 seconds on average with 32 attributes whereas the update and verification take even less, only 0.08 seconds on average. Whilst this data point does not consider other activities a real user may experience (e.g., application rendering), it provides a positive indication that this is an acceptable latency\cite{b14}.

\begin{figure}[h]
 \begin{center}
 \begin{tikzpicture}
    \begin{axis}[
         legend style={at={(0.00,0.66)},{font=\tiny},anchor=north west},
         width=\textwidth,
         width=6cm,
         height=4.6cm,
         ymin = 0,
         ylabel={Time (s)},
         xlabel={Number of attributes},
         symbolic x coords={8, 16, 24, 32},
         xtick=data
         ]
        
     \addplot[mark=*,blue,
     error bars/.cd,
     y dir=both,
     y explicit
     ] plot coordinates {
         (8, 0.9613733196) +- (0.08, 0.08)
         (16, 0.9395194197) +- (0.03, 0.03)
         (24, 0.9758505964) +- (0.04, 0.04)
         (32, 0.9885176802) +- (0.06, 0.06)
     };
     \addlegendentry{$\mathtt{Hash_{PCH}}$}

     \addplot[smooth,color=red,mark=x,error bars/.cd,
     y dir=both,
     y explicit]
         plot coordinates {
         (8, 0.07281450272) +- (0.01, 0.01)
         (16, 0.07538673401) +- (0.02, 0.02)
         (24, 0.07413482189) +- (0.005, 0.005)
         (32, 0.07364800453) +- (0.001, 0.001)
         };
     \addlegendentry{$\mathtt{Update_{PCH}}$}
    
     \addplot[smooth,color=brown,mark=x,error bars/.cd,
     y dir=both,
     y explicit]
         plot coordinates {
         (8, 0.08984062195) +- (0.01, 0.01)
         (16, 0.08984062195) +- (0.01, 0.01)
         (24, 0.08984062195) +- (0.01, 0.01)
         (32, 0.08984062195) +- (0.01, 0.01)
         };
     \addlegendentry{$\mathtt{Verify_{PCH}}$}
     \end{axis}
     \end{tikzpicture}
     \end{center}
 \vspace{-0.5em}
 \caption{Execution time of Template Hashing Algorithms}
 \label{fig:ExecTime}
 \end{figure}
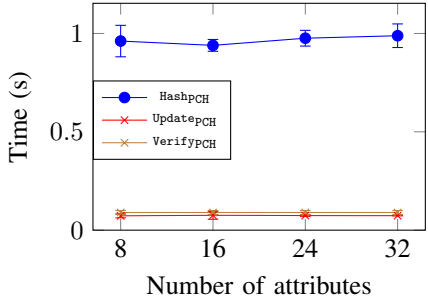

\begin{figure*}[h]
\small

\begin{subfigure}{0.47\textwidth}
\begin{tikzpicture}
\begin{axis}[
        legend style={at={(0.50,0.66)},{font=\tiny},anchor=north west},
        width=\textwidth,
        ymin = 0,
        width=9cm,
        height=4.6cm,
        xlabel={Concurrent hash generation requests},
        symbolic x coords={0, 50, 100, 150, 200, 250, 300, 350, 400, 450, 500},
        xtick=data
        ]

    \addplot[smooth,color=red,mark=x,error bars/.cd,
    y dir=both,
    y explicit]
        plot coordinates {
        (0, 0)
        (50, 0.494)
        (100, 0.38)
        (150, 0.388)
        (200, 0.416)
        (250, 0.52)
        (300, 0.315)
        (350, 0.362)
        (400, 0.361)
        (450, 0.35)
        (500, 0.355)
        };
    \addlegendentry{Average response time(s)}
    
    \addplot[smooth,color=blue,mark=x,error bars/.cd,
    y dir=both,
    y explicit]
        plot coordinates {
        (0, 0)
        (50, 36.167)
        (100, 35.712)
        (150, 35.857)
        (200, 35.922)
        (250, 36.167)
        (300, 35.877)
        (350, 35.912)
        (400, 35.829)
        (450, 36.025)
        (500, 35.959)
        };
    \addlegendentry{Average block transfer(KB/s)}

     \addplot[smooth,color=brown,mark=x,error bars/.cd,
    y dir=both,
    y explicit]
        plot coordinates {
        (0, 0)
        (50, 7)
        (100, 8)
        (150, 7)
        (200, 7)
        (250, 7)
        (300, 8)
        (350, 8)
        (400, 8)
        (450, 8)
        (500, 7)
        };
    \addlegendentry{Requests Processed per Second(RPS)}
    
\end{axis}
\end{tikzpicture}
    \vspace{-0.5em}
\caption{Throughput (Template hashing and signing)}
\label{fig:TPS}
\end{subfigure}
\begin{subfigure}{0.47\textwidth}
\begin{tikzpicture}
        \begin{axis}[
        legend style={at={(0.50,0.32)},{font=\tiny},anchor=north west},
        width=\textwidth,
        ymin = 0,
        width=9cm,
        height=4.6cm,
        xlabel={Concurrent hash update requests},
        symbolic x coords={0, 50, 100, 150, 200, 250, 300, 350, 400, 450, 500},
        xtick=data
        ]

    \addplot[smooth,color=red,mark=x,error bars/.cd,
    y dir=both,
    y explicit]
        plot coordinates {
        (0, 0)
        (50, 10.6332)
        (100, 17.3628)
        (150, 26.0611)
        (200, 33.3617)
        (250, 25.6860)
        (300, 25.2219)
        (350, 25.8208)
        (400, 28.4664)
        (450, 22.6828)
        (500, 23.0452)
        };
    \addlegendentry{Average response time(ms)}
    
    \addplot[smooth,color=blue,mark=x,error bars/.cd,
    y dir=both,
    y explicit]
        plot coordinates {
        (0, 0)
        (50, 36.170)
        (100, 36.170)
        (150, 35.006)
        (200, 32.629)
        (250, 31.317)
        (300, 30.233)
        (350, 31.506)
        (400, 27.066)
        (450, 29.266)
        (500, 30.976)
        };
    \addlegendentry{Average Block Transfer(KB/s)}

     \addplot[smooth,color=brown,mark=x,error bars/.cd,
    y dir=both,
    y explicit]
        plot coordinates {
        (0, 0)
        (50, 12.4)
        (100, 12.2)
        (150, 12.6)
        (200, 13.1)
        (250, 14.2)
        (300, 14.8)
        (350, 13.8)
        (400, 14.9)
        (450, 15.0)
        (500, 14.2)
        };
    \addlegendentry{Requests Processed per Second(RPS)}
    
\end{axis}
    \end{tikzpicture}
    \vspace{-0.5em}
    \caption{Throughput(Template update)}
\label{fig:TPS_update}
\end{subfigure}
    \vspace{-0.3em}
\caption{System load testing}
\label{fig:image2}
\end{figure*}
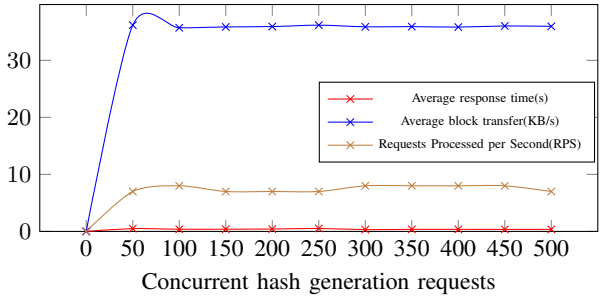
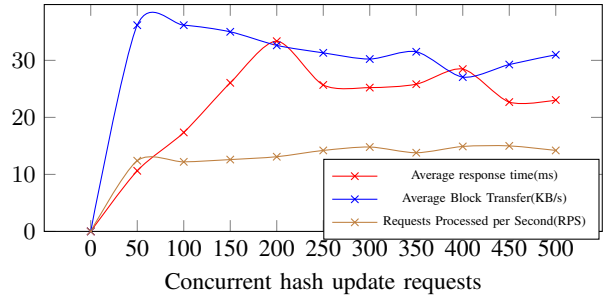

\begin{table}[ht!]
    \centering
    \scriptsize
    \caption{}
    \small
   \begin{tabular}{ |p{1.0cm}|| p{0.3cm} | p{0.3cm} | p{0.3cm} | p{0.3cm} | p{0.3cm} | p{0.3cm} | p{0.3cm}| p{0.3cm} | p{0.3cm} | p{0.3cm} |}
    \hline
    Users & 50 & 100 & 150 & 200 & 250 & 300 & 350 & 400 & 450 & 500\\
    \hline
    Requests & 34 & 79 & 116 & 147 & 15 & 124 & 282 & 213 & 254 & 346\\
    \hline
    \end{tabular}
    \label{tab:hash_gen}
\end{table}

\begin{table}[ht!]
    \centering
    \scriptsize
    \caption{}
    \small
   \begin{tabular}{ |p{1.0cm}|| p{0.3cm} | p{0.3cm} | p{0.3cm} | p{0.3cm} | p{0.3cm} | p{0.3cm} | p{0.3cm}| p{0.3cm} | p{0.3cm} | p{0.3cm} |}
    \hline
    Users & 50 & 100 & 150 & 200 & 250 & 300 & 350 & 400 & 450 & 500\\
    \hline
    Requests & 229 & 339 & 311 & 235 & 328 & 329 & 318 & 298 & 372 & 376\\
    \hline
    \end{tabular}
    \label{tab:hash_update}
\end{table}

\paragraph{\underline{Performance of Template Hashing and Signing}}  
 
For system load testing purposes, we created a test scenario based on the workflow depicted in Fig.~\ref{fig:test}. During the test, we increase the number of users from 50 up to 500 sending concurrent template generation requests. We also varied the number of requests from each user group to simulate varying system loads. This varied system load is generated independently by our testing tool. Table \ref{tab:hash_gen} shows the number of concurrent requests sent within each user group.

Fig.~\ref{fig:TPS} shows the test results. The average load of blocks transferred in the blockchain during the load test period is represented in Kilobytes (Kb/s). The average response time is represented in seconds. The plot in Fig.~\ref{fig:TPS} illustrates that the time taken increases proportionally with the number of concurrent users. To summarise the numbers on the graph, with 50 concurrent users accessing the system, we recorded 36.167 KB/s of average block transfer time, an average response time of 0.494s and 7 requests processed per second (RPS). With 500 users, we recorded 35.959 KB/s of average block transfer time, an average response time of 0.355s and 7 requests/s are processed. We can observe that with varying system loads, the average response time and average block transfer time remained stable.

\paragraph{\underline{Performance of Template In-Use (Updates)}} 

To measure time taken to update a template, we tested Step~2 from Fig.~\ref{fig:test} (2a + 2b). Step~3 are identical to Step~2. We used the same load testing scenario as described above. The number of
concurrent requests sent within each user group is shown in Table \ref{tab:hash_update}. 

Fig.~\ref{fig:TPS_update} shows the test results. To summarise the numbers on the graph, with 50 concurrent users, we recorded 36.170KB/s of average block transfer time, an average response time of 10ms and 12.4 requests processed per second (RPS). With 500 users, we recorded 30.976KB/s of average block transfer time, an average response time of 23ms and 14.2 requests/s are processed. We can observe that with varying system loads, the average response time and average block transfer time show a decreasing trend, although the average block transfer time kept increasing after 400 users, i.e., going from 27KB/s to 30KB/s. 

Compared to the template hashing and signing phase, this template in-use phase exhibits a lower average response time, while a higher number of requests are processed per second. The possible reason for this result could be attributed to the lower execution time of the template update algorithm ($\mathtt{Update_{PCH}}$) as shown in Fig.~\ref{fig:ExecTime}.

 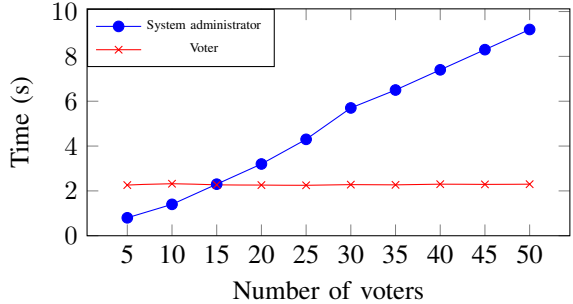
\begin{figure}[h]
 \begin{center}
 \begin{tikzpicture}
    \begin{axis}[
         legend style={at={(0.00,1.0)},{font=\tiny},anchor=north west},
         width=\textwidth,
         width=8cm,
         height=4.6cm,
         ymin = 0,
         ylabel={Time (s)},
         xlabel={Number of voters},
         symbolic x coords={5,10,15,20,25,30,35,40,45,50},
         xtick=data
         ]
        
     \addplot[mark=*,blue,
     error bars/.cd,
     y dir=both,
     y explicit
     ] plot coordinates {
         (5, 0.8)
         (10, 1.4)
         (15, 2.3)
         (20, 3.2)
         (25, 4.3)
         (30, 5.7)
         (35, 6.5)
         (40, 7.4)
         (45, 8.3)
         (50, 9.2)         
     };
     \addlegendentry{System administrator}

     \addplot[smooth,color=red,mark=x,error bars/.cd,
     y dir=both,
     y explicit]
         plot coordinates {
         (5, 2.26)
         (10, 2.32)
         (15, 2.27)
         (20, 2.26)
         (25, 2.25)
         (30, 2.28)
         (35, 2.27)
         (40, 2.3)
         (45, 2.29)
         (50, 2.3)  
         };
     \addlegendentry{Voter}

     \end{axis}
     \end{tikzpicture}
     \end{center}
 \vspace{-0.5em}
 \caption{The average time cost for the system administrator and the voter based on the number of voters participating in the voting process}
 \label{fig:voting_cost}
 \end{figure}

\paragraph{\underline{Performance of Voting Process}} 
To see how the time cost for the system administrator and the voter vary with an increasing number of voters we have carried out experiments with 5 to 50 voters. Each voter needs to {\it{review}} the update in a credential (e.g., adding Anna as a personal issuer) and then to {\it{vote}} confirming the update. Fig.~\ref{fig:voting_cost} highlights the distribution of time cost for the system administrator and the voter based on the number of voters participating in the voting process. Overall, running the voting process with 50 voters costs the system administrator 9.2s. The total time including the time for the administrator and the voters is 124.2s which breaks down to a reasonable time of 2.3s per voter to {\it{review}} and {\it{vote}}. This performance analysis shows that the system administrator’s time cost increases linearly based on the number of voters (to verify role credential and vote count), and the voter’s time cost remains constant.

\subsection{Security Analysis of VCTP Protocol}

In this section, we consider security threats of the VCTP protocol by considering potential attack scenarios in distributed trust management systems\cite{b15} and show that the proposed protocol can guard against these attacks.

\textbf{Attack scenario 1 - Secret information intercepts.} In this scenario, an attacker has intercepted the hashed template, encrypted trapdoors  shared with an updater via a secure communication channel. An attacker may be attempt to make an illegitimate update on the template.

\textbf{Response:} First, the sharing of secret information is restricted and limited in our system. However, to mitigate the potential risk of the communication being intercepted, the hashed template and encrypted trapdoors can be further protected by encrypting them with the receiver's (i.e., updater) public key so that only the intended recipient can decrypt the shared information. This can additionally ensure the receiver and signer accountability.

\textbf{Attack scenario 2 - Impersonation.} In this scenario, Issuer A may attempt to update a section of a credential authorised for Issuer B, by impersonating Issuer B. This may be possible if the trapdoor keys have the same access attributes for updatable sections.

\textbf{Response:} In our system, each trapdoor key is encrypted with a different set of attributes, requiring the updater to possess attributes that are appropriate for the context of the section being updated. Managing access to the trapdoor keys is a critical security aspect of the protocol, and the designer of the template should carefully consider the attributes to minimise overlaps. 

\textbf{Attack scenario 3 - Collusion.} In this scenario, a trust proxy may collude with an external user and try to onboard him/her as a personal issuer. That is, for instance, a doctor may try and onboard someone who is not a patient. The malicious trust proxy may attempt this by updating a section to specify the external user as the personal issuer. Then, the malicious personal issuer may try and submit fake attributes to access the attribute key generator in an attempt to receive the attribute private key. 

\textbf{Response:} For a successful authorisation, the submitted attributes must be valid and provable from a verifiable credential. In addition, our proposed template is specific to a particular domain (e.g., our template example is specific to healthcare domain) and any user who does not own the designated access attributes of this domain cannot be onboarded as a personal issuer. In this way, collision is further limited to a particular domain. Moreover, each trust proxy would have to complete the voting process satisfactorily before propagating his template to an external user. These demonstrate that our protocol significantly minimises the possibility of collusion among the participating entities.

\begin{table*}[ht]
    \centering
    \caption{Blockchain based Issuer Management Systems for Trust}
    \begin{tabular}{ | p{3cm} | p{3cm} | p{3cm} | p{3cm} | p{3cm} | }
    \hline
    Schemes & SSI context & level of issuers & update control & non-linear verification \\
    \hline
    CredTrust \cite{b6} & \checkmark & \checkmark & $\times$ & $\times$ \\
    \hline
    Grüner et. al. \cite{b16} & \checkmark & $\times$ & $\times$ & $\times$\\
    \hline
    Manas et. al. \cite{b17} & \checkmark & $\times$ & $\times$ & $\times$ \\
    \hline
    Schlatt et. al.\cite{b18}  & \checkmark & \checkmark & $\times$ & $\times$  \\
    \hline
    Soltani et. al.\cite{b2}  & \checkmark & \checkmark & $\times$ & $\times$ \\
    \hline
    Sovrin\cite{b19}  & \checkmark & \checkmark & $\times$ & $\times$ \\
    \hline
    Nandakumar et. al. \cite{b20} & $\times$ & $\times$ & \checkmark & \checkmark \\
    \hline
    Pal et. al.\cite{b21} & $\times$ & \checkmark & $\times$ & $\times$ \\
    \hline
    Proposed VCTP & \checkmark & \checkmark & \checkmark & \checkmark \\
    \hline
    \multicolumn{5}{l}{$\times$ indicates that the scheme does not consider the feature or has not published technical details.}
    \end{tabular}  
    \label{tab:RelatedWork}
\end{table*}

\section{Related Works}
\label{sec:related-work}

In this section, we summarise the studies on the topic of trust management for blockchain-based issuers. We organise the studies into two categories: solutions with SSI, and those without SSI.

\paragraph{\underline{Issuer management in SSI systems}} 
Authors in \cite{b6} constructed an issuer trust management process for SSI-based ecosystem. The trust management process supports the idea of a chain of issuers where issuers with a higher level of trust (e.g., government agencies) can propagate trust to the next level of issuers (i.e., non-officiated/individual issuers). This work presents a credential-based trust propagation and verification process along with verifiable credential templates which encode and validate trust propagation rules. However, the work fails to avoid collusion among the entities.

In \cite{b16}, authors propose a quantifiable trust model for the digital identities in blockchain-based SSI. The trust model uses directed graphs to determine the trust flow between identities and calculates a quantitative trust value for a claim (i.e., a plain statement about a digital identity) by the number of attestations (i.e., a statement from an entity to assert the correctness of a claim) it has received. It further derives overall trust values for identities from the trust scores of claims issued to identities. Although this work presents the trust value for all identities in SSI, it did not provide any issuer credential or systematic solution specifying how an issuer can gather their trust scores claiming him as an issuer.

Manas et. al. \cite{b17} propose a quantitative model for computing reputation scores for SSI issuers. The average number of credentials issued over a certain period is considered a factor to calculate the reputation score. This reputation score is further combined with the identity trust score from \cite{b16} to determine a quantitative confidence level for an issuer. An issuer with a higher confidence level allows peers to freely do future transactions without security or privacy concerns. Lower confidence enables the peers to exchange information cautiously, or even break the peer relationship.

The authors in \cite{b18} designed a framework for ``Know-Your-Customer'' (KYC) processes that manage client identities for banking applications. The proposed KYC architecture is built on blockchain-based SSI to afford owners (i.e., bank customers) better control on their data (e.g., owners can get a permanent overview of whom they shared what data with). Soltani et. al. \cite{b2} were the first to explore SSI in the context of KYC, covering the onboarding process for issuers and customers and technically evaluating their solution using Hyperledger Indy (public permissioned DLT). The Indy architecture supports various built-in roles such as trustee, trust anchor, and identity owner. The Trustee who has the highest privileges in the system registers the trust anchor (e.g., a bank) in the ledger as an officiated issuer. Officiated issuers can issue identity credentials to their customers who in turn can use these credentials with other banks as proof of identity. Another similar issuer onboarding idea proposed in \cite{b22} considers only the officiated issuers as authorised enlisting in the blockchain network. 

Sovrin is an open-source SSI project that offers the tools and libraries to run an identity management system on its network. Its identity system defines multiple roles for governance authorities who can onboard "Trust Anchors" (e.g., banks, or hospitals) as the official issuers. These "Trust Anchors" can onboard other issuers in the Sovrin network. Although, Sovrin allows an individual issuer to self-issue a credential (e.g., Alice can self-certify that she likes chocolate), there are some reasonable boundaries on self-issuance\cite{b23}. For instance, Alice cannot self-certify that she is nominating her sister as a guardian because she is recently suffering from dementia. For such a certification, Sovrin has defined an identity control system called guardianship where a guardian (i.e., Alice's sister) represents a dependent (i.e., Alice) \cite{b19, b24}. The guardianship identity control system is given a credential schema. The issuers of this credential are managed by a pre-installed governance authority. For example, guardianship management is governed by ``jurisdiction'' which defines objectives, stakeholders and duties/rights in a guardianship. 

Our proposed VCTP protocol aims to establish a trust layer for issuers so that the verifiers and holders can build trust relationships with them. This trust layer allows an individual (e.g., Anna) to be onboarded as a trusted personal issuer so that when needed they can pass their rights to other users (e.g., Tim).

\paragraph{\underline{Issuer management in non-SSI systems}} 

Nandakumar et. al. \cite{b20} has used a sanitizable signature to allow an authorised issuer to update the multimedia content before the content is distributed. Here blockchain is considered to maintain an immutable log of the updated segments to establish the provenance and integrity of multimedia data. Unlike the proposed VCTP protocol, this paper considered only the content owner (i.e., the issuer) as the authorised future updater to update the content.

Pal et. al. \cite{b21} has proposed a trust propagation method in the IoT context for delegating access right using blockchain. For the verification of access delegation, a smart contract is proposed to follow the delegation trail on-chain until the source of root delegation is found. The search time for verifying each level of delegation is the limiting factor for such a system. 

\paragraph{\underline{Use of sanitizable signature in different context}} 

Sanitizable signature is used in mutable blockchain to update the stored data on-chain when needed\cite{b5}. Nandakumar et. al. \cite{b20} has utilised this signature to update stored multimedia content prior to public sharing. However, none of them has considered sanitizable signature to update credential policies. In this paper, we are proposing to apply sanitizable signature to update credential policies as a way of propagating trust to the credential issuers.

We compare the proposed VCTP protocol with other blockchain-based issuer management systems, along the key properties of the architecture, namely: SSI context, a chain of issuers, update control, and non-linear verification. Table \ref{tab:RelatedWork} summarises the comparison. In summary, the papers that considered the level of issuers in their management system neither ensure updated control on propagating trust to the personal issuers nor avoid linear verification while verifying the issuer signature.

\section{Conclusion}
\label{sec:conclusion}

In this paper, we proposed the VCTP protocol to support trust propagation to an individual issuer. The levels of issuers and credential based trust propagation allow individuals to be onboarded as personal issuers with credibility. In addition, we have proposed a credential template for the issuers to allow for a secure and increased applicability of SSI in real-world scenarios. To minimise the possibility of collusion among the participating entities, we have taken measures in a form of the voting process. Finally, the feasibility of the proposed design is demonstrated through the proof of concept implementation of the VCTP protocol. A more in-depth assessment of the proposed protocol in real-world setup along with an efficient credential revocation process paves the path for future endeavors.

\section*{Acknowledgment}

This work is supported through "Australian Government Research Training Program" funded by "Commonwealth Department of Education and Training". Hye-young Paik has been partly supported by Cyber Security Collaborative Research Centre, Australia.


\end{document}